\documentclass{article}
\usepackage{ijcai22}

\usepackage{amssymb,mathrsfs,amsmath}
\usepackage{times}
\usepackage{soul}
\usepackage{url}
\usepackage[hidelinks]{hyperref}
\usepackage[utf8]{inputenc}
\usepackage[small]{caption}
\usepackage{graphicx}
\usepackage{amsmath}
\usepackage{amsthm}
\usepackage{booktabs}
\usepackage{algorithm}
\usepackage{algorithmic}
\usepackage{microtype}
\usepackage{graphicx}
\usepackage{subfigure}
\usepackage{balance}
\usepackage{booktabs} 
\usepackage{amsthm}
\usepackage{array}
\usepackage{graphicx}
\usepackage{clrscode}
\usepackage{subfigure}
\usepackage{multirow}
\usepackage{multicol}
\usepackage{float}
\usepackage{color}
\usepackage{xcolor}
\usepackage{amsopn}
\usepackage{mathrsfs}
\usepackage{mathtools}
\usepackage{amsmath}
\usepackage{booktabs}
\usepackage{arydshln}
\usepackage{hyperref}
\usepackage{blkarray}
\usepackage{enumerate}
\usepackage{courier}
\usepackage{mathrsfs}
\usepackage{rotating}
\usepackage{bm}
\usepackage{subfigure}
\usepackage{array}
\usepackage{ragged2e}
\usepackage{hyperref}
\usepackage{amsmath}

\theoremstyle{definition}

\theoremstyle{theorem}

\theoremstyle{proof}

\theoremstyle{remark}
\newtheorem*{remark}{Remark}

\renewcommand\arraystretch{1.2}

\title{Measuring ``Why'' in Recommender Systems: a Comprehensive Survey on the Evaluation of Explainable Recommendation}

\author{
Xu Chen$^1$
\and
Yongfeng Zhang$^2$
\and
Ji-Rong Wen$^1$
\affiliations
$^1$Beijing Key Laboratory of Big Data Management and Analysis Methods,\\
Gaoling School of Artificial Intelligence, Renmin University of China\\
$^2$Department of Computer Science, Rutgers University
\emails
xu.chen@ruc.edu.cn,
yongfeng.zhang@rutgers.edu,
jrwen@ruc.edu.cn
}

\begin{document}
\maketitle

\begin{abstract}
Explainable recommendation has shown its great advantages for improving recommendation persuasiveness, user satisfaction, system transparency and among others. 
A fundamental problem of explainable recommendation is how to evaluate the explanations.
In the past few years, various evaluation strategies have been proposed.
However, they are scattered in different papers, and there lacks a systematic and detailed comparison between them.
To bridge this gap, in this paper, we comprehensively review the previous work, and provide different taxonomies according to the evaluation perspectives and evaluation methods.
Beyond summarization, we also analyze the (dis)advantages of existing evaluation methods, and provide a series of guidelines on how to select them.
The contents of this survey are concluded from more than 100 papers from top-tier conferences like IJCAI, AAAI, TheWebConf, Recsys, UMAP and IUI, and the complete comparisons are presented at https://shimo.im/sheets/VKrpYTcwVH6KXgdy/MO\\DOC/.
With this survey, we finally aim to provide a clear and comprehensive review on the evaluation of explainable recommendation.
\end{abstract}

\section{Introduction}
Artificial intelligence (AI) has deeply revolutionized people's daily life and production, bringing effectiveness, efficiency, reliability, automation and etc.
Among different AI tasks, many are purely objective, where the ground truths are irrelevant with subjective factors.
Computer vision (CV) is a typical domain falling into this category.
For example, in the task of image classification, the models are built to accurately predict the labels like a cat or a dog.
These labels are objective facts, which do not change with human feelings.
In many other domains, the tasks are more subjective, where there is no rigid ground truth, and the target is to improve the utilities of certain stakeholders.
Recommender system is a typical subjective AI task, where the stakeholder utilities not only include the recommendation accuracy, but also contain a lot of beyond-accuracy aspects, among which explainability is a significant and widely studied one.

Basically, explainable recommendation solves the problem of ``why'', that is, why the items are recommended.
By providing explanations, people can make more informed decisions, and the recommendation persuasiveness and algorithm transparency can be improved.
In the field of explainable recommendation, evaluation is a fundamental problem, which judges the model availability and helps to make model selections.
Comparing with the other evaluation tasks, evaluating recommendation explanations is much more difficult because the ground truth is hard to obtain and human feelings are not easy to approximate.
To alleviate these difficulties, the past few decades have witnessed a lot of promising evaluation strategies.
However, these strategies are independently proposed in different papers, and has various prerequisites and implementation characters.
There lacks a systematic comparison between them, which may hinder the evaluation standardization and be not friendly to the newcomers to this domain.

To bridge the above gap, in this paper, we thoroughly conclude the previous work, and present a clear summarization on existing evaluation strategies.
In specific, recommendation explanations may serve for different targets, for example, the users, providers and model designers.
For different targets, the evaluation perspectives are also various, and we summarize four major evaluation perspectives from the previous work, that is, the explanation effectiveness, transparency, persuasiveness and scrutability. 
For evaluating the explanations from the above perspectives, there exists a lot of evaluation methods.
We conclude these methods into four categories including the case studies, quantitative metrics, crowdsourcing and online experiments.
For each method, we detail its main characters, and analysis its advantages and shortcomings, 
At last, we propose a series of guidelines on how to select these evaluation methods in real-world problems.
The main contributions of this paper are summarized as follows:

$\bullet$ We provide a systematic and comprehensive survey on the evaluation of explainable recommendation, which to the best of our knowledge, is the first survey on this topic.

$\bullet$ We propose different taxonomies for existing papers according to the evaluation perspectives and evaluation methods.
The references cover more than 100 papers from top tier conferences like IJCAI, AAAI, TheWebConf, Recsys, UMAP and IUI, and their complete summarization is presented at https://shimo.im/sheets/VKrpYTcwVH6KXgdy/MODOC/.

$\bullet$ We analysis the advantages and shortcomings of existing evaluation methods and propose several guidelines on how to choose them in different scenarios.


\section{Related Survey}
Recommender system is becoming increasingly important, with dozens of papers published each years on top tier conferences/journals.
For better summarizing and comparing these work, people have conducted many promising survey studies.
For general recommendation, early surveys mainly focus on shallow architectures like matrix factorization or heuristic methods~\cite{adomavicius2004recommendation}.
Later on, with the ever prospering of deep neural networks, people have designed a large amount of neural recommender models.
Correspondingly, many comprehensive surveys on deep recommender algorithms are proposed~\cite{zhang2019deep,wu2021survey,fang2020deep}.
Beyond concluding recommender models from the architecture perspective, many surveys focus on how to leverage the side information.
For example, ~\cite{bouraga2014knowledge} summarizes the models on incorporating knowledge graph into recommender systems.
~\cite{chen2013social} reviews the algorithms on social recommendation.
~\cite{kanwal2021review} focuses on the combination between the text information and recommender models.
~\cite{ji2017survey} summarizes the models on image based recommendation.
In addition, people also conduct surveys according to different recommendation domains.
For example, ~\cite{song2012survey} summarizes the algorithms in music recommendation.
~\cite{goyani2020review} focus more on the movie recommender models.
~\cite{wei2007survey} is tailored for the algorithms in e-commerce.
~\cite{yu2015survey} is a comprehensive survey on point-of-interest (POI) recommendation.
We have also noticed that there are some surveys on recommendation evaluations~\cite{gunawardana2009survey,silveira2019good}.
However, they focus more on the metrics like accuracy, novelty, serendipity and coverage, which differs from our target on explanation evaluation.
In the past, we have published a survey on explainable recommender models~\cite{zhang2018explainable}, but it aims to summarize this field in a general sense.
However, in this paper, we focus on the evaluation part, and include many latest studies.

\begin{table*}[t]
\centering
\caption{Summarization of the evaluation perspectives.}
\vspace*{-0.2cm}
\scalebox{1.}{
\begin{tabular}
{
m{3.2cm}<{\centering}|
m{5.cm}<{\centering}|
m{4.5cm}<{\centering}|
m{3.cm}<{\centering}}
\hline \hline
Evaluation perspective  &Evaluation problem &Representative papers&Serving target\\ \hline
Effectiveness&Whether the explanations are useful for the users to make more accurate/faster decisions?&~\cite{hada2021rexplug,guesmi2021demand,wang2020personalized,gao2019explainable,chen2013tagcloud}&Users\\  \hline
Transparency&Whether the explanations can reveal the internal working principles of the recommender models?&~\cite{chen2021temporal,sonboli2021fairness,li2021you,li2020directional,fu2020fairness}&Model designers\\ \hline
Persuasiveness&Whether the explanations can increase the click/purchase rate of the users on the items?&~\cite{musto2019justifying,tsai2019evaluating,balog2020measuring,tsukuda2020explainable,chen2014sentiment}&Providers\\ \hline
Scrutability&Whether the explanations can exactly correspond to the recommendation results?&~\cite{he2015trirank,tsai2019evaluating,cheng2019incorporating,balog2019transparent,liu2020explainable}&Model designers\\ \hline \hline
\end{tabular}
}
\label{compare1}
\vspace{-0.3cm}
\end{table*}

\section{Evaluation Perspectives}\label{ep}
Recommendation explanations may serve for different purposes, \emph{e.g.}, providing more recommendation details, so that people can make more informed decisions,
or improving recommendation persuasiveness, so that the providers can obtain more profit.
For different purposes, the explanations should be evaluated from different perspectives.
In the literature, the following evaluation perspectives are usually considered.

\subsection{Effectiveness}
Effectiveness aims to evaluate ``whether the explanations are useful for the users to make more accurate/faster decisions~\cite{balog2020measuring}?'' or sometimes ``whether the users are more satisfied with the explanations~\cite{xian2021ex3}?''
It is targeted at improving the utilities of the users, that is, better serving the users with the explanations.
By enhancing the effectiveness, the user experience can be improved, which may consequently increase the user conversion rate and stickiness.
Here, we use the term ``effectiveness'' in a broad sense, which covers the concepts of efficiency, satisfaction and trust in many literature~\cite{balog2020measuring}.

\subsection{Transparency}
Transparency aims to evaluate ``whether the explanations can reveal the internal working principles of the recommender models~\cite{tai2021user,sonboli2021fairness,li2021attribute}?''
It encourages the algorithms to produce explanations which can open the ``black box'' of the recommender models.
With better transparency, the model designers can know more about how the recommendations are generated, and better debug the system.

\subsection{Persuasiveness}
Persuasiveness aims to evaluate ``whether the explanations can increase the interaction probability of the users on the items~\cite{tsukuda2020explainable,balog2020measuring,musto2019justifying}?''
In general, persuasiveness aims to enhance the utilities of the providers (\emph{e.g.}, sellers on the e-commerce websites, content creators on the micro video applications).
Comparing with transparency, persuasiveness does not care about whether the explanations can honestly reflect the model principles.
It focuses more on whether the explanations can persuade users to interact with the recommendations, so that the business goals can be achieved.

\subsection{Scrutability}
Scrutability aims to evaluate ``whether the explanations can exactly correspond to the recommendation results?''~\cite{cheng2019incorporating,tsai2019evaluating,he2015trirank}.
In other words, if the explanations are different, then whether and how the recommendation list is altered.
This perspective is closely related with transparency yet one does not imply the other.
Transparency pays more attention to the details inside the model, aiming to whiten the working mechanisms.
Scrutability cares more about the relations between the explanations and outputs, which can be model agnostic.

We summarize the above evaluation perspectives in Table~\ref{compare1}.
We can see, recommendation explanations mainly serve for three types of stakeholders, that is, the users, providers and model designers.
For different stakeholders, the evaluation perspectives are different: effectiveness cares more about the utilities of the users, persuasiveness is more tailored for the providers, and transparency and scrutability serve more for the model designers.
If one compares the explanations in recommender systems and general machine learning, she may find that the latter mainly focus on the transparency and scrutability, since it is not related with the users and provides. 
In this sense, explainable recommendation can be more complex, since it needs to trade-off different stakeholders.

\begin{remark}
(1) The above four perspectives are not always independent, they may have both overlaps and contradictions.
For example, the explanations for improving transparency may satisfy the curiosities of some users on how the recommender models work, and therefore the explanation effectiveness can be enhanced. 
In another example, to persuade users to purchase more items (\emph{i.e.}, {persuasiveness}), the explanations may not accurately reveal the model working mechanism (\emph{i.e.}, {transparency}), but have to highlight the features which can promote sales.
When designing and evaluating explainable recommender models, it is necessary to predetermine which perspectives to focus on, and how much we can sacrifice on the other perspectives.
(2) We have noticed there are many other evaluation perspectives, such as the fluency of the explanation~\cite{chen2021towards} or whether a user-item pair is explainable~\cite{liu2019in2rec}. However, these perspectives are not common, only appearing in a small amount of papers. Thus we do not elaborate them in this survey.
\end{remark}

\section{Evaluation Methods}
To evaluate recommendation explanations from the above perspectives, people have designed quite a lot of evaluation methods.
By going over the previous papers, we find that there are mainly four categories, that is: case studies, quantitative metrics, crowdsourcing and online experiments.
In the following, we elaborate these methods more in detail.

\subsection{Evaluation with Case Studies}
In this category, people usually present some examples to illustrate how the recommender model works and whether the generated explanations are aligned with human intuitions.
Usually, the recommender models are firstly learned based on the training set, and then the examples are generated based on the intermediate or final outputs by feeding the testing samples into the optimized model.
More specifically, the examples can be both positive and negative~\cite{chen2019personalized}.
By the positive examples, the effectiveness of the model can be demonstrated, while by the negative ones, the readers can understand when and why the model fails.

The most popular case studies are conducted by visualizing the attention weights~\cite{li2021attribute,li2021you,chen2019personalized}.
For example, in fashion recommendation, ~\cite{li2021attribute} presents the matching degree between the tops and bottoms across different attributes.
In image recommendation, ~\cite{chen2017attentive} visualizes the importances of different user previously interacted images as well as the regions in each image.
In graph based recommendation, ~\cite{li2021you} illustrates which neighbors, attributes and structure contexts are more important for the current user-item interaction prediction.

In many studies, people present template-based or fully generated natural language explanations as case studies.
For example, ~\cite{chen2020try,zhang2014explicit} show the explanations by assembling the model predicted attributes and the pre-defined templates.
~\cite{li2020towards,li2021personalized} present the explanations which are completely generated from the models.

When evaluating knowledge-based recommender models, people usually conduct case studies by showing the reasoning paths between the users and items.
For example, 
in~\cite{wang2019explainable}, the authors present the importances of different paths for a given user-item pair.
In~\cite{xian2020cafe,fu2020fairness}, the authors present examples to reveal the model routing process from the users to the items.

In many papers, the explanation items/features are presented as intuitive case studies.
For example,~\cite{liu2020explainable,li2020directional} highlights the items which are significant for the current model prediction.
~\cite{yang2019interpretable} shows important feature combinations to reveal the model working principles.

In aspect-based recommender models, the case studies are usually conducted by presenting the learned aspects for the users/items.
For example, in~\cite{tan2016rating}, the authors show the top words in the learned aspects to demonstrate the effectiveness of the model in capturing user preference. 

The major advantage of case studies lies in its intuitiveness, which makes the readers easily understand how the explanations look like.
However, there are many significant weaknesses.
To begin with, it can be biased, since one cannot evaluate and present all the samples.
And then, due to the lack of quantitative scores, it is difficult to compare different models and make accurate model selections.

\subsection{Evaluation with Quantitative Metrics}
In order to more precisely evaluate the explanations, people have designed many quantitative metrics.
Basically, this category of methods assumes that the stakeholder utilities can be approximated and evaluated via parameterized functions.
In the following, we elaborate the representative metrics.

Intuitively, if the explanations can accurately reflect the user preferences on the items, then the user satisfaction can be improved.
As a result, many papers~\cite{li2021personalized,hada2021rexplug,li2020generate} leverage the user reviews, which pool comprehensive user preferences, as the explanation ground truth.
In these studies, the explainable recommendation problem is regarded as a natural language generation (NLG) task, where the goal is to accurately predict the user reviews.
Under this formulation, the following metrics are usually leveraged to evaluate the explanation qualities.

$\bullet$ \textbf{BLEU and ROUGE scores.}
BLEU~\cite{papineni2002bleu} and ROUGE~\cite{lin2004rouge} scores are commonly used metrics for natural language generation, where the key insight is to count the world (or n-gram) level overlaps between the model predicted and real user reviews.

$\bullet$ \textbf{Unique Sentence Ratio (USR), Feature Coverage Ratio (FCR) and Feature Diversity (FD).}
These metrics aim to evaluate the diversity of the generated reviews~\cite{li2020generate}.
USR evaluates the diversity on the sentence level.
In specific, suppose the set of unique reviews is $\mathcal{S}$, and the total number of reviews is N, then this metric is computed as $\text{USR} = \frac{|\mathcal{S}|}{N}$.
FCR and FD compute the diversity on the feature level.
In the former metric, suppose the number of distinct features in all the generated reviews is $M$, and the total number of features is $N_f$, than $\text{FCR} = \frac{M}{N_f}$.
For the latter one, suppose the feature set in the generated review for user-item pair $(u,i)$ is $F_{u,i}$, then $\text{FD} = \frac{1}{N(N-1)}\sum_{(u,i)\neq (u',i')} |F_{u,i}\cap F_{u',i'}|$, where $N$ is the total number of different user-item pairs.

$\bullet$ \textbf{Feature-level Precision (FP), Recall (FR) and $F_1$ (FF)}
In many papers~\cite{tan2021counterfactual,tai2021user}, the explanation quality is evaluated by comparing the features in the predicted and real user reviews.
For each user-item pair $(u,i)$, suppose the predicted and real feature sets are $S_{u,i}$ and $T_{u,i}$, respectively.
Then the feature-level precision, recall and $F_1$ for this user-item pair are computed as $\text{FP}_{u,i} = \frac{|S_{u,i}\cap T_{u,i}|}{|S_{u,i}|}$, $\text{FR}_{u,i} = \frac{|S_{u,i}\cap T_{u,i}|}{|T_{u,i}|}$ and $\text{FF}_{u,i} = \frac{2\cdot \text{FP}\cdot \text{FR}}{\text{FP} + \text{FR}}$, respectively. 
The final results are averaged across all the user-item pairs, that is, $\text{FP} = \frac{1}{N}\sum_{u,i}\text{FP}_{u,i}$, $\text{FR} = \frac{1}{N}\sum_{u,i}\text{FR}_{u,i}$ and $\text{FF} = \frac{1}{N}\sum_{u,i}\text{FF}_{u,i}$, where $N$ is the number of user-item pairs.

$\bullet$ \textbf{Feature Matching Ratio (FMR).} 
In~\cite{li2020generate}, the reviews are generated under the control of a feature.
FMR evaluates whether this feature can successfully appear in the predicted results.
It is formally computed as $\text{FMR} = \frac{1}{N}\sum_{u,i}\bm{1}(f_{u,i}\in S_{u,i})$, where $f_{u,i}$ is the input feature, $S_{u,i}$ is the predicted review, $N$ is the number of user-item pairs, and $\bm{1}$ is the indicator function.

In counterfactual explainable recommendation~\cite{tan2021counterfactual}, the following two metrics are usually leveraged to evaluate how the model prediction changes when the key features/items (which are used as the explanations) are altered.

$\bullet$ \textbf{Probability of Necessity (PN).}
This metric aims to evaluate ``if the explanation features of an item had been ignored, whether this item will be removed from the recommendation list?''
Formally, Suppose $A_{ij}$ is the explanation feature set of item $j$ when being recommended to user $i$, let $\text{PN}_{ij}=1$, if item $j$ no longer exists in the recommendation list when one ignores the features in $A_{ij}$, otherwise $\text{PN}_{ij}=0$.
The final PN score is computed as $\sum_{i,j} \frac{\text{PN}_{ij}}{\bm{1}({|A_{ij}|>0})}$.

$\bullet$ \textbf{Probability of Sufficiency (PS).}
This metric aims to evaluate ``if only the explanation features of an item are remained, whether this item will be still in the recommendation list?''
Let $\text{PS}_{ij}=1$, if item $j$ exists in the recommendation list when one only remains the features in $A_{ij}$, otherwise $\text{PS}_{ij}=0$.
The final PS score is computed as $\sum_{i,j} \frac{\text{PS}_{ij}}{\bm{1}({|A_{ij}|>0})}$.

Similar evaluation metric is also leveraged in~\cite{liu2020explainable}, which measures the explanation quality based on the performance change before and after the key history items/entities are removed from the model input.
The key intuition is that if the explanations are the key history items/entities underlying the recommendations, then removing them should greatly change the recommendation performance. 
The formal definition is as follows:

$\bullet$ \textbf{Performance Shift (PS).}
Suppose the original recommendation performance is $r$, after removing the key history items/entities, the performance is changed to $r'$, then $\text{RS} = \frac{r-r'}{r}$.
Here the performance can be evaluated with different metrics, such as precision, recall, NDCG and so on.

In~\cite{abdollahi2017using}, the authors firstly define which items can be explained based on the user-item interaction graph, and then the following two metrics are proposed to evaluate the explainability of the recommendations.

$\bullet$ \textbf{Mean Explainability Precision (MEP) and Mean Explainability Recall (MER).}
For a user $u$, suppose the set of items which can be explained is $N_u$, the recommended item set is $M_u$, then $\text{MEP} = \frac{|N_u\cap M_u|}{|M_u|}$ and $\text{MEP} = \frac{|N_u\cap M_u|}{|N_u|}$.

Based on the above quantitative metrics, different models can be compared, which is important for model selection and benchmarking this field.
In addition, quantitative metrics are usually very efficient, since most of them can be computed in a short time.
As for the shortcomings, quantitative metrics are sometimes not fully aligned with the evaluation goals. 
For example, BLUE score only evaluates the word-level overlaps between the generated and real explanations, but it does not directly tell whether the explanations are reasonable or not.
What's more, when the stakeholder utilities are too complicated, existing quantitative metrics, which are heuristically designed by a small amount of people, may have the risk of misleading the evaluation results.

\subsection{Evaluation with Crowdsourcing}
As mentioned before, recommender system is a subjective AI task, thus involving human feelings into the evaluation process is a direct and natural idea.
Such evaluation methods are called crowdsourcing, and there are mainly three strategies, which are elaborated as follows.

$\bullet$ \textbf{Crowdsourcing with public datasets.}
In this method, the recommender models are trained based on the public datasets.
And then, many annotators are recruited to evaluate the model generated explanations based on a series of questions.
There are two key points in the evaluation process.

(\romannumeral1) Annotation quality control.
In order to accurately evaluate the explanations, controlling the annotation quality is a necessary step.
In the previous work, there are two major strategies:
one is based the voting mechanism, for example, in~\cite{chen2019personalized}, there are three annotators for each labeling task, and the final result is available only when more than two annotators have the same judgment.
The other strategy is based on computing certain statistical quantities.
For example, in~\cite{chen2018neural}, Cohen’s Weighted $\kappa$~\cite{cohen1968weighted} is leveraged to assess the inter-annotator agreements and remove the outlier annotations.

(\romannumeral2) Annotation question designs.
We find that the annotation questions in different papers are quite diverse.
In general, they can be classified into three categories.
The first category are ``model-agnostic'' questions, where the problems do not dependent on any specific model.
In~\cite{chen2019personalized}, the annotators are required to simulate themselves as the real users, and then label out the ideal explanations from all the possible ones.
Then the labeled explanations are regarded as the ground truth, and the proposed model and baselines are evaluated by comparing their predicted explanations with the ground truth.
The second category are ``single-model'' questions, where the problems focus on only one explainable model.
In~\cite{li2021attribute}, the annotators are asked to up-vote or down-vote the explanations generated from the designed model and the baseline, respectively.
In~\cite{xian2021ex3,li2020towards}, the annotators have to evaluate whether the proposed model can help users make better purchase decision.
In~\cite{chen2021towards}, the authors ask the annotators to label whether the explanations produced from the designed model are fluent and useful.
In~\cite{tao2019fact,wang2018explainable}, the annotators are required to answer whether they are satisfied with the generated explanations, whether the explanations can really match their preferences and whether the explanations can provide more information for users to make a decision.
The third category are ``pair-model'' questions, where the problems focus on the comparison between a pair of explainable models.
In~\cite{li2020towards}, the annotators need to answer which explanations produced from the designed model and baseline are more closer to the ground truth.
In~\cite{tao2019fact}, the annotation questions are: between A and B, whose explanations do you think can better help you understand the recommendations? and whose explanations can better help you make a more informed decision?
In~\cite{chen2018neural}, the annotators are asked to compare the usefulness of the explanations generated from the proposed model and the baselines.

In the above questions, the first category is the most difficult, since the annotators have to make decisions from a large amount of candidates. However, the labeled datasets are model-agnostic, which can be reused for different models and experiments.
In the second and third categories, the annotators only have to answer yes or no, or select from a small amount of candidate answers, but the annotations cannot be reused, which is costly and not scalable.

$\bullet$ \textbf{Crowdsourcing by injecting annotator data into public dataset.}
A major problem in the above method is that the annotator preferences may deviate from the real users', which may introduce noises into the evaluation.
As a remedy for this problem, ~\cite{gao2019explainable} designs a novel crowdsourcing method by combining the annotator generated data with the public dataset.
In specific, the evaluations are conducted based on the Yelp dataset\footnote{https://www.yelp.com/dataset}, which is collected from yelp.com.
The authors recruit 20 annotators, and require each annotator to write at least 15 reviews on yelp.com.
Then, the reviews written by the annotators are infused into the original Yelp dataset, and the explainable models are trained based on the augmented dataset.
In the evaluation process, the annotators are required to score from 1 to 5 on the explanations according to their usefulness.
Since the data generated by the annotators is also incorporated into the model training process, the feedback from the annotators are exactly the real user feedback.

\renewcommand\arraystretch{1.3}
\begin{table*}[t]
\centering
\caption{Summarization of the evaluation methods.
In the second and third column, we present the most significant strengths and shortcomings of different evaluation methods.
In the last column, we present the perspectives that a method is usually leveraged to evaluate.
}
\vspace*{-0.2cm}
\scalebox{1.}{
\begin{tabular}
{
m{3.cm}<{\centering}|
m{2.8cm}<{\centering}|
m{2.8cm}<{\centering}|
m{3.5cm}<{\centering}|
m{3.5cm}<{\centering}}
\hline \hline
Evaluation methods  &Strengths &Shortcomings&Representative papers&Evaluation perspectives\\ \hline
Case studies& Better intuitiveness & Bias; Cannot make comparisons &~\cite{xian2020cafe,xian2020cafe,fu2020fairness,liu2020explainable,barkan2020explainable,li2020directional}&Effectiveness; Transparency\\  \hline
Quantitative metrics& Quantitative evaluation; Easy to benchmark; High efficiency & Deviating from the explanation goals; Less effective approximation &~\cite{li2021personalized,hada2021rexplug,li2020generate,tan2021counterfactual,tai2021user}&Effectiveness; Scrutability\\  \hline
Crowdsourcing& Based on real human feelings & High cost &~\cite{li2021attribute,xian2021ex3,li2020towards,chen2021towards,wang2018explainable}&Effectiveness; Scrutability; Transparency; Persuasiveness\\  \hline
Online experiments& High reliable & High cost&~\cite{zhang2014explicit,xian2021ex3}&Effectiveness; Persuasiveness\\  \hline\hline
\end{tabular}
}
\label{compare2}
\vspace{-0.2cm}
\end{table*}

$\bullet$ \textbf{Crowdsourcing with fully constructed datasets.}
In this category, the datasets are fully constructed by the annotators.
In general, there are four steps in the evaluation process, that is, 
(\romannumeral1) recruiting annotators with different background (\emph{e.g.}, age, sex, nationality and etc.), 
(\romannumeral2) collecting annotator preferences, 
(\romannumeral3) generating recommendations and explanations for the annotators, 
and (\romannumeral4) evaluating the explanations by asking questions on different evaluation perspectives.
Usually, The number of annotators in this category is much larger than above two methods, and the experiment settings and studied problems are quite diverse.
For example, in~\cite{musto2019justifying}, there are 286 annotators (with 76.3\% males, and 48.4\% PhDs), and they are asked to evaluate the explanation transparency and effectiveness.
In~\cite{naveed2020use}, the authors aim to study whether feature-based collaborative filtering (CF) models can lead to better explanations than the conventional CF's.
There are 20 annotators, among which 14 are females with the age ranges from 21 to 40.
In~\cite{hernandez2020effects}, the authors investigate whether the type and justification level of the explanations influence the user perceptions.
To achieve this goal, 152 annotators are recruited, where there are 87 females, and the average age is 39.84.
In~\cite{balog2020measuring}, the authors recruit 240 annotators to study the relations between the explanation effectiveness, transparency, persuasiveness and scrutability.
In~\cite{tsukuda2020explainable}, the authors recruit 622 annotators to evaluate the influence of the explanation styles on the explanation persuasiveness.

In the above three types of crowdsourcing methods, the first one is fully based on the public datasets, the second one is based on the combination between the annotator generated data and the public dataset, and the last one is completely based on the annotator generated data.
Comparing with the case studies and quantitative metrics, crowdsourcing is directly based on real human feelings.
However, it can be much more expensive due to the cost of recruiting annotators.

\subsection{Evaluation with Online Experiments}
In the previous work, we also find a few papers, which evaluate recommendation explanations with online experiments.
In specific, in~\cite{zhang2014explicit}, the online users are split into three groups.
For the users in the first group, the explanations are generated from the proposed model.
For the second group users, the explanations are produced from the baseline.
For the last group, there is no explanations. 
After running the system for a short time, the authors compute the click through rate (CTR) in each group to evaluate whether the explanations are useful.
Similar method are also leveraged in~\cite{xian2021ex3}, where the conversion and revenue are compared before and after presenting the explanations to the users.
We believe online experiments can be more reliable. However, its cost is very high, since it may disturb the real production environments, and impact user experiences.

We summarize the above evaluation methods in Table~\ref{compare2}, where we can see:
for different evaluation methods, they are usually leveraged to evaluate different perspectives.
For example, case studies are mainly used to evaluate explanation effectiveness and transparency, while crowdsourcing can be leveraged to evaluate all the perspectives defined in section~\ref{ep}, since it can flexibly design questions for the annotators. 
From the second and third columns, we can see different evaluation methods have their own strengths and shortcomings, and no one method can take all the advantages.

\textbf{Guidelines for selecting the evaluation methods.}
Based on different characters of the evaluation methods, we propose several guidelines on using them in real-world applications.
(\romannumeral1) If the recommender models serve for high-stake tasks such as health caring and finance, then more reliable evaluation methods should be selected, since the cost of mis-explaining a recommendation, so that the users fail to make correct decisions can be much larger than that of conducting the evaluations.
(\romannumeral2) For general recommendation tasks, if the evaluation budget is limited, one has to trade-off the strengths and shortcomings of different evaluation methods.
(\romannumeral3) From Table~\ref{compare2}, we can see the same perspective can be evaluated using different methods, thus one can simultaneously using these methods to take their different advantages. 
For example, when the model is not well optimized, we can use case studies to roughly evaluate the explanation effectiveness.
Once the model has been tuned better, quantitative metrics can be leveraged to demonstrate whether the explanations are indeed effective.

\vspace{-0.1cm}
\section{Conclusion and Outlooks}
In this paper, we summarize existing explainable recommendation papers with a focus on the explanation evaluation.
In specific, we introduce the main evaluation perspectives and methods in the previous work.
For each evaluation method, we detail its characters, representative papers, and also highlight its strengths and shortcomings. 
At last, we propose several guidelines for better selecting the evaluation methods in real-world applications.

By this survey, we would like to provide a clear summarization on the evaluation strategies of explainable recommendation.
We believe there is still much room left for improvements.
To begin with, since the heuristically designed quantitative metrics can be incompetent for evaluating the complex stakeholder utilities, one can automatically learn the metric functions, where one can incorporate a small amount of human labeled data as the ground truth.
And then, for benchmarking this field, one can build a large scale dataset, which incorporates the ground truth on the user effectiveness, recommendation persuasiveness and so on.
At last, existing methods mostly evaluate the explanations within a specific task.
However, we believe human intrinsic preferences should be stable and robust across different tasks.
Thus, in order to evaluate whether the derived explanations can indeed reveal such human intrinsic preferences, developing task-agnostic evaluation methods should be an interesting research direction.

\bibliographystyle{named}
\bibliography{ijcai22}

\end{document}